\documentclass[twocolumn,aps,prl,10pt,showpacs,showkeys,]{revtex4}
\usepackage[dvips]{graphicx}

\begin{document}
\title{Crystal-field states in NaCoO$_{2}$ and in wet cobalt superconductors (Na$_{x}$CoO$_{2}$)$^\spadesuit$}
\author{R. J. Radwanski}
\homepage{http://www.css-physics.edu.pl}
\email{sfradwan@cyf-kr.edu.pl} \affiliation{Center of Solid State
Physics, S$^{nt}$Filip 5, 31-150 Krakow, Poland,\\
Institute of Physics, Pedagogical University, 30-084 Krakow,
Poland}
\author{Z. Ropka}
\affiliation{Center of Solid State Physics, S$^{nt}$Filip 5,
31-150 Krakow, Poland}

\begin{abstract}
We claim that for calculations of the electronic structure of 3d
oxides very strong electron correlations have to be taken into
account similarly to those assumed in many-electron crystal field
approach. For Co$^{3+}$ ions in Na$_{x}$CoO$_{2}$$\cdot$yH$_{2}$O
there are 15 low lying many-electron states within 0.1 eV as can
be obtained in the many-electron CEF approach with taking into
account the spin-orbit coupling. We claim, that the used by us
Hamiltonian for the trigonal distortion is correct.

\end{abstract}
\maketitle

Na$_{x}$CoO$_{2}$ compounds attract a broad interest after
discovery of superconductivity in their hydrated compounds \cite
{1}.

An aim of this paper is two fold. At first we would like to put
attention that the electronic structure of the trivalent cobalt
ion occurring in Na$_{x}$CoO$_{2}$ should be calculated with
taking into account very strong electron correlations. Secondly
we would like to clarify the description of the trigonal
distortion of the octahedral crystal field. A direct motivation
for this paper is a correction of erroneous electronic structure
presented in \cite {2}, where correlations have been taken to be
rather weak.

The cobalt ion sits in maternal Na$_{x}$CoO$_{2}$ compounds as
well as in their hydrades Na$_{x}$CoO$_{2}$$\cdot$yH$_{2}$O in a
local oxygen octahedron \cite {1,2}. The diagonal of the local
octahedron is along the hexagonal axis of the hexagonal unit
cell. In such structural construction a trigonal distortion can
be realized by extending/compression of the distance between the
whole oxygen planes. Exactly the same construction of the O-Co-O
planes occurs in the perovskite structure, in LaCoO$_{3}$ \cite
{3} for instance, perpendicularly to the cube diagonal, and even
in FeBr$_{2}$ \cite {4}, having a hexagonal unit cell. In
LaCoO$_{3}$ there exists Co$^{3+}$ ions, whereas in FeBr$_{2}$
the Fe$^{2+}$ ion is formed. Both these ions have six electrons
in the incomplete 3d shell forming, according to us, a
strongly-correlated atomic-like 3$d^{6}$ configuration. Their
low-energy electronic structure has been calculated by us taking
into account the translational symmetry characteristic for the
crystalline compound, very strong electron correlations,
intra-atomic spin-orbit coupling, multipolar lattice charge
interactions known as crystal-field interactions with a trigonal
distortion of the dominant octahedral crystal field \cite {3,4}.
For a relatively weak octahedral CEF the ground state is a state
originating from the high-spin $^{5}T_{2g}$ cubic subterm - such
a situation is realized in FeBr$_{2}$ and it prefers magnetic
ground state \cite {4}. In LaCoO$_{3}$ the octahedral CEF is
stronger, due to the doubly negative charge of the O ions
compared to the single valency of the Br ions and especially due
to the small Co-O distance of 193 pm, a low-spin $^{1}A_{1}$
($^{1}I$) cubic subterm (singlet!!) becomes the ground state
\cite {3}. In this case a non-magnetic state in the atomic scale
is formed.

In Na$_{x}$CoO$_{2}$ compounds both Co$^{3+}$ and Co$^{4+}$ ions
coexist. In this paper, for a clarity reason, we confine our
consideration to NaCoO$_{2}$ (x=1), where the Co$^{3+}$ ions only
exists and a problem of charge disproportionalization or
segregation does not occur. NaCoO$_{2}$ has been sintesized
recently \cite {5} and it has been found that the Co$^{3+}$ ions
in NaCoO$_{2}$ are in a non-magnetic state and that there is only
single Co site. Thus we think that the origin of this non-magnetic
state in NaCoO$_{2}$ is the same as in LaCoO$_{3}$ \cite {3},
where we took into account strong electron correlations and
crystal-field interactions. The high energy, 0-4 eV, electronic
structure associated with the splitting of single-ion terms by
octahedral crystal field is shown in Fig. 2. Depending on the
sodium and water content this non-magnetic state can be changed
to the magnetic one similarly to the 3$d^{6}$ ion in FeBr$_{2}$
because then the strength of CEF decreases leaving the
$^{5}$$T_{2g}$ subterm as the ground subterm. The splitting of
the $^{5}$$T_{2g}$ subterm is shown in Fig. 3.
\begin{figure}[t]
\begin{center}
\includegraphics[width = 6.6 cm]{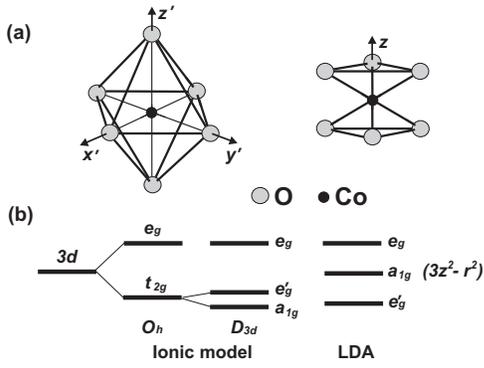}
\end{center} \vspace {-0.3 cm}
\caption{(a) Illustration of the trigonal distortion of a
CoO$_{6}$ octahedron. Left panel: undistorted CoO$_{6}$ octahedron
with cubic ($O_{h}$) symmetry. Right panel: compressed CoO$_{6}$
octahedron with $D_{3d}$ symmetry. The distorted CoO$_{6}$ is
rotated such that the threefold rotation axis is along the $c$
axis. (b) Crystal-field splitting of Co 3$d$ states in distorted
CoO$_{6}$ according to an ionic model and relative energy
positions of 3$d$ bands obtained from LDA calculations. Redrawn
from Ref. \cite{2} together with the full caption. This
electronic structure, calculated without electron correlations, we
consider to be very oversimplified \cite {11} - we claim that a
proper one for NaCoO$_{2}$ is that shown in Fig. 3.}
\end{figure}

\begin{figure}\vspace {-0.2 cm}
\begin{center}
\includegraphics[width = 7.0 cm]{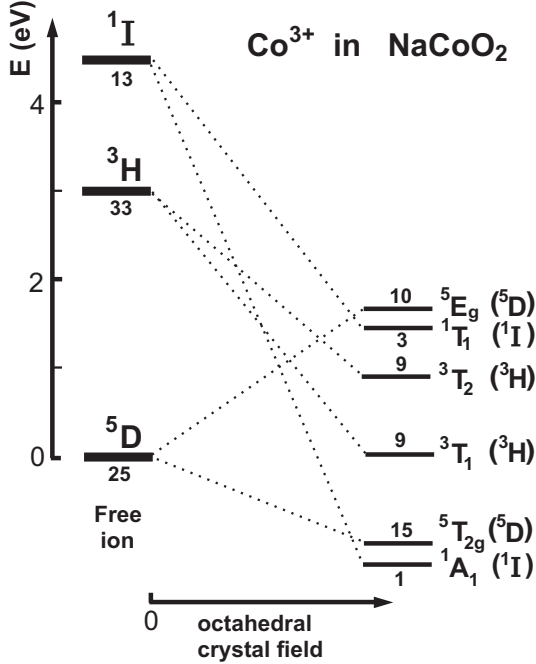}
\end{center} \vspace {-0.6 cm}
\caption{Electronic structure of cubic subterms of the Co$^{3+}$
ion in the octahedral crystal field inferred from the
Tanabe-Sugano diagram for $Dq$/$B$ =2.2 relevant to NaCoO$_{2}$.
\vspace {-0.4 cm}}
\end{figure}
\begin{figure}\vspace {-0.2 cm}
\begin{center}
\includegraphics[width = 5.9 cm]{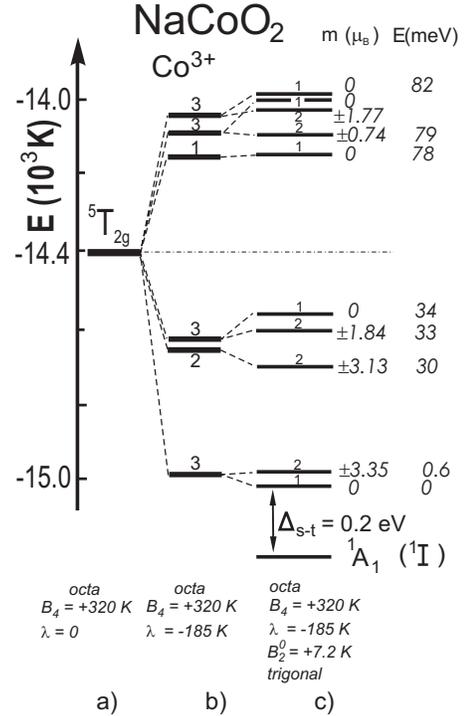}
\end{center} \vspace {-0.7 cm}
\caption{Calculated low-energy electronic structure of the
Co$^{3+}$ ion in
NaCoO$_{2}$ originating from the $^{5}T_{2g}$ cubic subterm with the $%
^{1}A_{1}$ singlet ground subterm put about 0.2 eV below the
lowest $^{5}T_{2g}$ state. Such the structure is produced by the
dominant octahedral crystal-field interactions and the
intra-atomic spin-orbit coupling (b). c) shows the splitting
produced by the trigonal compressed distortion. The states are
labelled by the degeneracy, the magnetic moment and the energy
with respect to the lowest state of the $^{5}D$ term. \vspace
{-0.7 cm}}
\end{figure}

We propagate the ionic model. We do not claim, however, to invent
the crystal-field theory, as it was invented in 1929-1932 by
Bethe, Kramers, Van Vleck and many others, but in last 20 years we
propagate the CEF approach being continuously discriminated
scientifically. This discrimination with some inquisition
incidents of e.g. Polish scientific institutions with a help of
Prof. H. Szymczak, J. Sznajd, J. Klamut, A. M. Oles, entitles us
to feel at present ourselves as reinventors of the crystal-field
in the solid-state physics \cite{6}, the more that we extend it
from the single-ion theory to the Quantum Atomistic Solid-State
theory, QUASST, pointing out for 3d-ion compounds, for instance,
the importance of the spin-orbit coupling \cite{7} and of local
distortions. We assume on-site electron correlations to be
sufficiently strong to keep the atomic-like integrity of the 3$d$
ion. QUASST deals consistently with both paramagnetic and
magnetically-ordered states.

For completeness we have to add that 25 states shown in our
papers [3,4] are only a small part of the full ionic electronic
structure of the Co$^{3+}$/Fe$^{2+}$ ion, that accounts in total
210 states grouped in 16 atomic terms \cite{8,9,10}.

We do not know reasons for forgetting works on the crystal field
of early Van Vleck, Tanabe and Sugano, and of many others. The
effect of the octahedral CEF interactions on these 16 terms have
been calculated by Tanabe and Sugano already 50 years ago
\cite{8,9}. These Tanabe-Sugano diagrams have been somehow
forgotten in the modern solid state theory \cite{6,7}, likely due
to an erroneous conviction that these states are not relevant to
solid materials. We are somehow grateful to the scientific
discrimination of e.g. Polish scientific institutions, to Prof.
Prof. H. Szymczak, J. Sznajd, J. Klamut, A. M. Oles because their
negative opinions are the best proof for the shortage of
knowledge about the (many-electron) crystal field in the XXI
century solid-state theory \cite {12}. One of reasons for the
limited use of the Tanabe-Sugano diagrams was caused by the large
uncertainty in the strength of the octahedral CEF in a particular
compound. From this point of view the exact evaluation of the
strength of the octahedral CEF in LaCoO$_{3}$ \cite {3} we
consider to be of the great importance.

Prof. H. Szymczak {\cite{13} has claimed that used by us
description of the trigonal distortion within the crystal-field
theory  is erroneous. His reproach was not corrected by following
referees J. Sznajd, K. Wysokinski, J. Klamut and A. M. Oles. Thus
we repeat here this description after our paper \cite{4}.

The 25 levels, originated from the $^{5}$D term, and their
eigenfunctions have been calculated by the direct diagonalization
of the Hamiltonian (1) within the $|LSL_{z}S_{z}\rangle $ base.
It takes a form:\vspace {-0.1 cm}
\begin{equation}
H_{d}=H_{cub}+\lambda L \cdot S + B_{0}^{2}O\,_{0}^{2}+\mu
_{B}(L+g_{s}S)\cdot B  \eqnum{1} \vspace {-0.1 cm}
\end{equation}
The separation of the crystal-electric-field (CEF) Hamiltonian
into the cubic and off-cubic part is made for the illustration
reason as the cubic crystal field is usually very predominant. In
the hexagonal unit cell of FeBr$_{2}$ the local cube diagonal
lies along the hexagonal {\it c} axis. The related distortion can
be described as the trigonal distortion of the local octahedron.
The cubic CEF Hamiltonian takes, for the {\it z} axis along the
cube diagonal, the form \vspace {-0.1 cm}
\begin{equation}
H_{cub}~~=~~-~\frac{2}{3} B_{4}\cdot
(O_{4}^{0}-20\sqrt{2}O_{4}^{3})\vspace {-0.1 cm}
\end{equation}
where $O\,_{m}^{n}$ are the Stevens operators. The last term in
Eq. (1) allows studies of the influence of the magnetic field.
For remembering, the octahedral CEF Hamiltonian with the $z$ axis
along the cube edge takes a form:\vspace {-0.1 cm}
\begin{equation}
H_{d}=B_{4}\cdot(O_{4}^{0}+5O_{4}^{4})\vspace {-0.1 cm}
\end{equation}
Before the end we summarize the treatment of NaCoO$_{2}$ within
the Quantum Atomistic Solid-State theory (QUASST). QUASST seems
to be a standard approach to insulating transition-metal oxides
but due to unknown reasons is highly discriminated in the modern
solid-state physics. As reasons for the discrimination an
oversimplicity is given as well as very-wide undefined statement
that the description of magnetic and electronic properties of
NiO, YTiO$_{3}$, CoO, BaVS$_{3}$, LaMnO$_{3}$, LaCoO$_{3}$ is not
suitable for publication being not of the broad interest to the
physics community. This argument cannot be treated seriously as
all these compounds, starting from NiO are continuously discussed
in Phys. Rev. B and Phys. Rev. Lett. with the starting point that
their magnetic and insulating properties are not yet understood.
In QUASST approach we treat the stechiometric NaCoO$_{2}$, with
the perfect intrinsic crystalographic structure and the single Co
site, as a charge transfer-insulator. During the formation of the
compound there occur electron transfer from Na and Co towards
oxygen forming Na$^{+}$Co$^{3+}$O$_{2}^{2-}$. These charges are
localized explaining the insulating ground state. Na$^{+}$ and
O$^{2-}$ ions have closed shells, whereas the Co$^{3+}$ ion has
six d electrons. The magnetism and the low-energy electronic
structure we attribute to the Co$^{3+}$ ions. According to
QUASST, formulated in times of the itinerant treatment of 3d
electrons, the six d electrons of the Co$^{3+}$ ion form strongly
correlated-electron atomic-like system keeping their atomic-like
3d$^{6}$ integrity, what means that we expect the intra-atomic
correlations and the resulting electron term structure to be
preserved. Such atomic-like system in a crystal experiences the
electrostatic crystal field due to all surrounded charges. This
CEF modifies the term structure in the well-known and controlled
way, Fig. 2, and substantially removes the degeneracy, Fig. 3.

Although we mention NaCoO$_{2}$ as a charge-transfer insulator it
is better characterized as a Mott insulator if one takes a
definition that the Mott insulator occur due to strong electron
correlations that prevent inter-site electron hopping. In our
picture the inter-site electron hopping, after the
charge-transfer during the formation of the compound, is not
allowed in NaCoO$_2$ by intra-atomic correlations which cause
that a formation of the 3d$^5$ (Co$^{4+}$) or 3d$^7$ (Co$^{2+}$)
configurations is energetically unprofitable. The definition of
the Mott insulator could be used in order to point out that a
given compound is the insulator despite having the open d or f
shell. This definition seems to be in a spirit of the original
Mott problem that NiO is the insulator having the open 3d shell,
i.e. in order to distinguish Na or Al, for instance, with open 3s
and 2p shells that are metals from NiO that is an insulator.
Then, the Mott and charge insulators distinguish between
insulating NiO and MgO. There could be a problem with the
classification of stechiometric TiO$_2$, for instance, which is
an oxide with the transition-metal atom but which in the ideal
case with Ti$^{4+}$ ions gave up all 3d electrons taking the
close-shell configuration. Thus we propose to use the name of
Mott insulator for insulating compounds containing
transition-metal atoms, in particular for oxides.

In conclusion, we claim that for calculations of the electronic
structure very strong electron correlations have to be taken into
account similarly to those assumed in the many-electron CEF
approach. We claim, that the used by us Hamiltonian for the
trigonal distortion is correct in contrary to the Szymczak's
reproach. We claim that the electronic structure of the Co$^{3+}$
ion in NaCoO$_{2}$ is much more complex than that considered in
Ref. \cite {2} and is close to that obtained for FeBr$_{2}$ and,
in particular, for LaCoO$_{3}$. The triangular trilayer structural
blocks in NaCoO$_{2}$ \cite {14} are similar to those realized in
FeBr$_{2}$. The one-electron CEF structure shown in Fig. 1 from
Ref. \cite{2} is only applicable for one d electron in the
trigonally compressed octahedron but not to the Co$^{3+}$ ion
which has six d electrons. We claim that the many-electron
strongly-correlated CEF approach \cite {15}, the basis for
QUASST, is physically adequate for 3d oxides. This controversy
between one-electron CEF, many-electron CEF (QUASST) approach and
band approaches can be experimentally solved by observation, or
not, of the predicted electronic structure. From this point of
view it is interesting to find why LDA calculations provide
another ground state, i.e. $e_{g}^{'}$ as shown in Fig. 1b right
\cite {16,17}, than the single-ion CEF. The comparison would be
easier if band-structure results contain data verifiable
experimentally. At least, effective charges at relevant atoms and
resultant experimental predictions should be reported. The QUASST
calculations allow for {\it ab initio} calculations of magnetic
and electronic properties. We assume atomistic construction of
the matter in a sense, that the transition-metal atom becoming
the full part of a solid preserve largely its atomic-like
integrity. The potentials used in the crystal-field Hamiltonian
have clear physical meaning and can be calculated from {\it first
principles}, from the point charge model in the simplest version.
Despite of the difference in $a_{1g}$ or $e{_g}^{'}$ ground state
between the single-electron CEF and band theory the basic
difference is related to an itinerant treatment of 3d electrons
\cite {16,17} and to the origin of the $t_{2g}$ splitting. In the
band theory the energy splitting is determined by the kinetic
energy of electrons \cite {18} in contrary to the crystal-field
view that the splitting is due to lattice distortions completed
in QUASST with the spin-orbit coupling and the very detailed
charge distribution in the whole compound. We note, with
pleasure, the changing of the band-theory calculations towards
local effects and discrete energy states characteristic for the
crystal-field theory. The similar origin of the non-magnetic
state in NaCoO$_2$ will occur also in isostructural LiCoO$_2$,
where the similar oxygen octahedron occurs and the Co-O distance
is below 193 pm.

$^\spadesuit$ dedicated to Hans Bethe, Kramers and John H. Van
Vleck, pioneers of the crystal-field theory, to the 75$^{th}$
anniversary of the crystal-field theory, and to the Pope John
Paul II, a man of the freedom and the honesty in the human life
and in Science.


\vspace {-0.2 cm}
\begin{thebibliography}{9}
\bibitem{1} K. Takada, H. Sakurai, E. Takayama-Muromachi, F. Izumi, R. A.
Dilanian, and T. Sasaki, Nature (London) {\bf 422}, 53 (2003).

\bibitem{2} W. B. Wu, D. J. Huang, J. Okamoto, A. Tanaka, H. -J. Lin, F. C. Chou, A. Fujimori, and C. T.
Chen, Phys. Rev. Lett. {\bf 94,} 146402 (2005).

\bibitem{3} Z. Ropka and R. J. Radwanski, Phys. Rev. B {\bf 67}, 172401
(2003).

\bibitem{4} Z. Ropka, R. Michalski, and R. J. Radwanski, Phys. Rev. B {\bf
63}, 172404 (2001).

\bibitem{5} G. Lang, J. Bobroff, H. Alloul, P. Mendels, N. Blanchard,
and G. Collin, Phys. Rev. B {\bf 72}, 094404 (2005); cond-mat/0505668.

\bibitem{6} R. J. Radwanski and Z. Ropka, cond-mat/0504199 (2005).

\bibitem{7} R. J. Radwanski and Z. Ropka, cond-mat/9907140 (1999).

\bibitem{8} Y. Tanabe and S. Sugano, J. Phys. Soc. Japan {\bf 9}, 753
(1954).

\bibitem{9} S. Sugano, Y. Tanabe, and H. Kamimura {\it Multiplets of
Transition-Metal Ions in Crystals} (Academic Press, New York)
1970.

\bibitem{10} A. Abragam and B. Bleaney, {\it Electron
Paramagnetic Resonance of Transition Ions} (Clarendon Press,
Oxford) 1970, ch. 7.

\bibitem{11} R. J. Radwanski and Z. Ropka, cond-mat/0506615 (2005).

\bibitem{12} H. Szymczak has blaimed us in November 2001 that we,
using 2.1 eV for 10$Dq$, assume too weak CEF interactions - now
he can see that at present much smaller values, even of 0.5 eV
only, are used in theoretical band calculations of Ref. 2. We
suppose that he writes a Comment to Phys. Rev. Lett. that too weak
crystal field is erroreously considered in Ref. \cite {2}, note
32. In Phys. Rev. Lett. and Phys. Rev. B can be found many such
papers.

\bibitem{13} Prof. H. Szymczak from the Polish Academy of Sciences, the
present long-lasting chairman of the Physical Department of PAS,
has claimed in an official administrative opinion in November
2001 for the Highest Scientific Council of the Polish Govermnent
that the procedure used by us and described in Ref. {\cite 4} is
incorrect. Prof. Szymczak wrote "Nawiasem mowiac dla Hamiltonianu
pola krystalicznego o symetrii oktaedrycznej zapisanego w postaci
B$_{4}$ (O$_{4}$$^{0}$ + 5O$_{4}$$^{4}$) wyraz
B$_{2}$$^{0}$O$_{2}$$^{0}$ opisuje deformacje tetragonalna a nie
trygonalna jak twierdzi dr hab. Ryszard J. Radwanski w swoich
publikacjach." what translated means: "Notabene, for a
crystal-field Hamiltonian with the octahedral symmetry written in
a form B$_{4}$ (O$_{4}$$^{0}$ + 5O$_{4}$$^{4}$) the term
B$_{2}$$^{0}$O$_{2}$$^{0}$ describes the tetragonal deformation,
but not the trigonal one {\bf as dr hab. Ryszard J. Radwanski
claims in his publications.}" This reproach is erroneous and
unfounded. Nowhere and never I claimed the supposition ascribed
to me by H. Szymczak as one can see, for instance, in Ref. {\cite
4}, the last acapit of the first page, a paper of mine written in
2000. Szymczak wrote his opinion in November 2001 having this
publication as the first position of my publication list. We
mention this problem here because if so great physicist as Prof.
H. Szymczak is does not know the CEF description of the trigonal
distortion it means that the knowledge about CEF is rather small.
Up to today, despite of my many requests for the withdrawing this
reproach, Prof. H. Szymczak did not correct this opinion. In
normal scientific conditions such a misunderstanding should be
easily solved - by Prof. Szymczak and other Polish professors it
becomes a way for the scientific inquisition. Another subject of
controversy was related to LaCoO$_{3}$ and to the ground state of
the Mn$^{3+}$ ion in LaMnO$_{3}$. Our paper with calculated by us
the ground subterm $^{5}E_{g}$, with excited the $^{5}T_{2g}$
states, was rejected at the SCES-02 Conference as being in
disagreement with the generally accepted view of the lowest
$t_{2g}$ states and higher $e_{g}$ states. We believe that
Science, Physics in particular, has to be made in the friendly
and well-wishing atmosphere. We can disagree about different
theoretical approaches but scientific problems have to be solved
in the open scientific discussion.

\bibitem{14} L. M. Helme, A. T. Boothroyd, R. Coldea, D. Prabhakaran,
A. Stunault, G. J. McIntyre, and N. Kernavanois, Phys. Rev. B {\bf
73}, 054405 (2006); cond-mat/0510360 (2005).

\bibitem{15} R. J. Radwanski and Z. Ropka, cond-mat/0404713
(2005).

\bibitem{16} D. J. Singh, Phys. Rev. B {\bf 61}, 13397
(2000); {\bf 68}, 020503(R) (2003).

\bibitem{17} K. -W. Lee, J. Kunes, and W. E. Pickett, Phys. Rev. B {\bf 70}, 045104
(2004).

\bibitem{18} W. Koshibae and S. Maekawa, Phys. Rev. Lett. {\bf 91}, 257003
(2003).

\end{thebibliography}
\end{document}